\documentclass[a4paper,10pt,doublespace]{article}
\usepackage{graphicx}
\usepackage{theorem}
\usepackage{enumerate}
\usepackage{amsmath}
\usepackage{amssymb}
\usepackage{a4wide}
\usepackage{url}

\newcommand{\lv}{\left \vert}
\newcommand{\rv}{\right \vert}
\newcommand{\la}{\left \langle}
\newcommand{\ra}{\right \rangle}
\newcommand{\ket}[1]{\lv #1 \ra}
\newcommand{\bra}[1]{\la #1 \rv}
\newcommand{\braket}[2]{\langle #1 \vert #2 \rangle}
\newcommand{\qed}{\hfill\hbox{\rule[-2pt]{3pt}{6pt}}\par}

\theorembodyfont{\rmfamily}
\newtheorem{Theorem}{\textit{Theorem} }

\newtheorem{Definition}{\textit{Definition} }
\newtheorem{Corollary}{\textit{Corollary}}
\newtheorem{Lemma}{\textit{Lemma}}
\newtheorem{Proof}{\textit{Proof}}

\begin{document}

\title{Exact Non-identity check is NQP-complete}
\author{Yu Tanaka\\
{\it Advanced Materials Laboratories, Sony}}
\date{\today}

\maketitle

\begin{abstract}
We define a problem ``exact non-identity check'': Given a classical description of a quantum circuit with an ancilla system, determine whether it is strictly equivalent to the identity or not. We show that this problem is NQP-complete. In a sense of the strict equivalence condition, this problem is different from a QMA-complete problem, non-identity check defined in \cite{identity}. As corollaries, it is derived that exact equivalence check is also NQP-complete and that it is hard to minimize quantum resources of a given quantum gate array without changing an implemented unitary operation.
\end{abstract}

\section{Introduction}

Non-identity check, which is defined and proven to be Quantum Merlin-Arthur (QMA) complete in \cite{identity}, is the problem of determining whether a given quantum circuit (unitary) is almost equivalent to a complex multiple of the identity with respect to the operator norm.
QMA is known to be a quantum analog of NP because it is an extension of the verifier-based definition of NP, in that sense non-identity check is one of the hardest problems based on quantum gate array complexity.

There is another quantum analog of NP, non-deterministic quantum polynomial-time (NQP)\cite{NQPDef}.
NQP is defined as an extension of the Machine-definition of NP, which is a set of decision problems solvable in polynomial time by a non-deterministic Turing machine. Here, we have a question: is there any NQP-complete problem based on quantum gate array complexity?

In this paper, to answer the question, we propose exact non-identity check which is the problem of determining whether a given classical description of a quantum circuit with an ancilla system is strictly equivalent to a complex multiple of the identity or not, and we prove that exact non-identity check is NQP-complete.

It is important to derive a NQP-complete problem based on quantum gate array complexity for at least two reasons. 
One is that such problems are less well-known because NQP does not equal QMA.
While two definitions of NP are equivalent, the following relation between QMA and NQP is known to be satisfied\cite{NQMA},
\begin{Lemma}\label{NQPQMA}
\begin{eqnarray}
NQP = \bigcup_{\delta : \mathbb{Z}^+ \to (0,1]} QMA(\delta, 0),
\end{eqnarray}
\end{Lemma}  
where QMA($\delta, 0$) is defined as QMA with perfect soundness in preliminary. 
Lemma \ref{NQPQMA} suggests that NQP does not equal QMA.
We use Lemma \ref{NQPQMA} to prove the NQP-hardness of exact non-identity check.
The other reason is that the NQP-complete problem is useful to analyze quantum gate array complexity. 
For example, we can show trivially that exact equivalence check is NQP-complete from exact non-identity check, where exact equivalence check is the problem of determining whether two unitary operations implemented by two given classical descriptions are strictly  equivalent or not.  
Further, it is derived to be hard to minimize quantum gate resources of a given quantum gate array without changing the content of the implemented unitary operation.

In this paper, first, we give our notations and definitions in preliminary. 
Second, we propose exact non-identity check and exact equivalence check, and prove the NQP-completeness of exact non-identity check and exact equivalence check. Further, quantum gates minimization problem is defined and proven to be NQP-hard.  
Finally, we summarize the exact non-identity check.

\section{Preliminary}

We start with giving several notations and definitions used in this paper.
A classical bit string $x \in \{ 0,1 \}^{\ast}$ is regarded as an integer if desired.
$\mathbb{Z}^+$ denotes the set of nonnegative integers.
$\mathcal{B} = \mathbb{C}^2$ denotes one qubit Hilbert space. For any Hilbert space $\mathcal{H}$, $\mathcal{S}(\mathcal{H})$ denotes the set of density operators over $\mathcal{H}$. $\mathcal{H}_{a}$ denotes Hilbert space for an ancilla system. Further, we define $\ket{\bar{0}} := \ket{0\cdots 0}$ and $\ket{x_+} := H^{\otimes n}\ket{x}$ for $x \in \{ 0,1 \}^n$, where $H$ is the Hadamard gate.

In order to use a result of Ref.\cite{NQMA}, we define ($\delta, \mu$)-quantum Merlin-Arthur (QMA).
The complexity class QMA or BQNP is a quantum analog of NP and was first defined in Ref.\cite{Kitaev}.

\begin{Definition} ({\bf QMA}($\delta, \mu$))\\
Given functions $\delta, \mu : \mathbb{Z}^+ \to [0,1]$, a language $L$ is in QMA($\delta, \mu$) if for every classical input $x \in \{ 0,1\}^{\ast}$ one can efficiently generate a quantum circuit $U_x$ (``verifier'') consisting of at most $p(|x|)$ elementary gates for an appropriate polynomial $p$ such that $U_x$ acts on the Hilbert space
\begin{eqnarray}
\mathcal{H} := \mathcal{B}^{\otimes n_x} \otimes \mathcal{B}^{\otimes m_x},
\end{eqnarray}
where $n_x, m_x$ grow at most polynomially in $|x|$. The first part is the input register and the second is the ancilla register.
Further, $U_x$ has the properties that 
\begin{eqnarray}
(\textrm{Completeness}) &&\forall x \in L \ \exists \rho \in \mathcal{S}(\mathcal{B}^{\otimes n_x}),\ {\rm Tr}[U_x (\rho \otimes \ket{\bar{0}}\bra{\bar{0}}) U_x^{\dagger}P_1] \ge \delta(|x|), \\
(\textrm{Soundness})&& \forall x \not\in L \ \forall \rho \in \mathcal{S}(\mathcal{B}^{\otimes n_x}),\ {\rm Tr}[U_x (\rho \otimes \ket{\bar{0}}\bra{\bar{0}}) U_x^{\dagger}P_1] \le \mu(|x|),
\end{eqnarray}
where $P_1$ is the projection corresponding to the measurement ``Is the first qubit in state 1?''. 
A quantum state $\rho$ is called a proof or a witness.
\end{Definition}

Note that a proof which is a mixed state does not increase the completeness due to the linearity of the quantum operation and the tracing out operation. Here, QMA with perfect soundness or NQMA\cite{NQMA} is defined.

\begin{Definition} ({\bf QMA with perfect soundness, NQMA})\\
A language $L$ is in NQMA if there exists a function $\delta : \mathbb{Z}^+ \to (0, 1]$ such that $L$ is in QMA($\delta$,0).
\end{Definition}

The complexity class NQP is a quantum analog of NP and was proposed as the class of the problems that are solvable in polynomial time by non-deterministic quantum Turing machines.

\begin{Definition} ({\bf NQP})\\
A language $L$ is in NQP if and only if there exists a quantum Turing machine $Q$ and a polynomial $p$ such that 
\begin{eqnarray}
\forall x \in L \Longleftrightarrow Pr[Q\ \textrm{accepts}\ x\ \textrm{in}\ p(|x|)\ \textrm{steps}] \neq 0.
\end{eqnarray}
\end{Definition}

So far, we can use the result in Ref.\cite{NQMA}, Lemma \ref{NQPQMA}, to prove the NQP-hardness of exact non-identity check.

\section{Exact non-identity check and exact equivalence check}

Exact non-identity check is the problem of determining whether a classical description of unitary operation is \textit{strictly  equivalent} to identity or not. Non-identity check proposed in Ref.\cite{identity} is, however, whether a given quantum circuit is the identity with respect to the operator norm or not.  
To state exact non-identity check problem precisely, we have to define an implemented unitary operation with an ancilla system. 

\begin{Definition}({\bf An implemented unitary operation with an ancilla system})\\
For every classical input $x \in \{ 0,1 \}^{\ast}$ one can efficiently generate a quantum circuit $U_x$ consisting of at most $p(|x|)$ elementary gates for an appropriate polynomial $p$ such that $U_x$ acts on the Hilbert space
$
\mathcal{H}_{in} \otimes \mathcal{H}_{a} := \mathcal{B}^{\otimes n_x} \otimes \mathcal{B}^{\otimes m_x},
$
where $n_x$ and $m_x$ grow at most polynomially in $|x|$.
The quantum circuit $U_x$ implements a unitary operation $U$ with an ancilla if $U_x$ satisfies that 
\begin{eqnarray}
\exists \ket{\phi_x} \in \mathcal{H}_{a}\ \forall \ket{\psi} \in \mathcal{H}_{in},\ U_x( \ket{\psi} \otimes \ket{\bar{0}}) = U\ket{\psi} \otimes \ket{\phi_x}. \label{implement}
\end{eqnarray} 
\end{Definition}

\begin{figure}
\begin{center}
\includegraphics[scale=0.7]{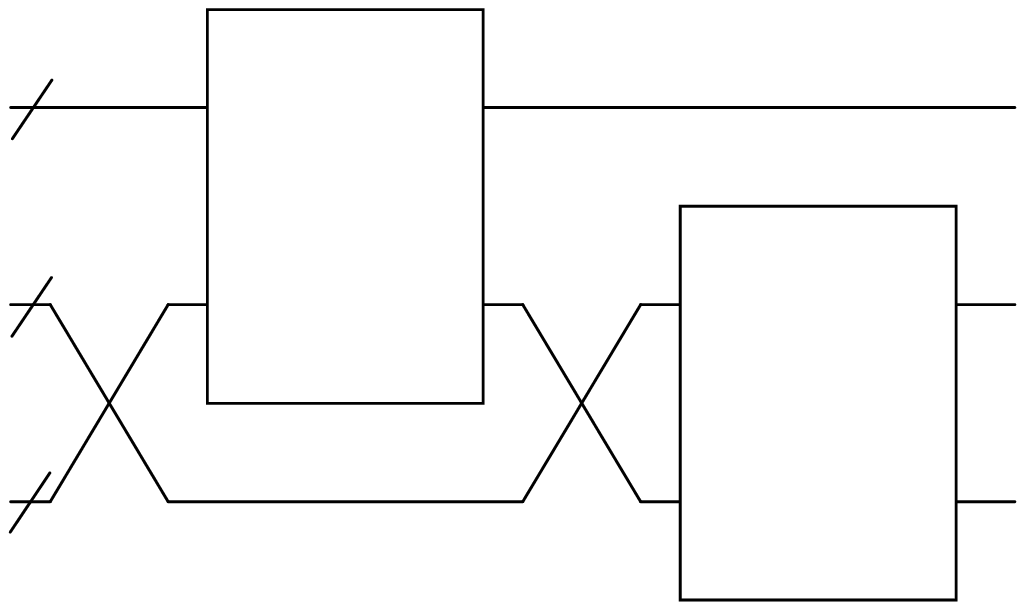}
\put(-245,102){Inputs}
\put(-245,62){Inputs$^\prime$}
\put(-248,22){Ancillas}
\put(-207,112){$n$}
\put(-207,73){$n$}
\put(-208,33){$m$}
\put(-147,80){\Large{$U_x$}}
\put(-51,40){\Large{$U_x^{\dagger}$}}
\caption{Circuit $Z_x$ consisting of $U_x$ and its complex conjugate.}
\label{FigZ00}
\end{center}
\end{figure}

In general, $\ket{\phi_x}$ is unknown. However, for a quantum circuit $U_x$ satisfying Eq.(\ref{implement}), we can always take $\ket{\phi_x} = \ket{\bar{0}}$ by constructing another quantum circuit $Z_x$ implementing $U \otimes U^{\dagger}$ in Fig. \ref{FigZ00}.
In the following, note that for a given classical description of $U_x$, $Z_x$ denotes the circuit in Fig. \ref{FigZ00}.

To define exact non-identity check, it is useful to show another equivalent representation of Eq.(\ref{implement}).

\begin{Lemma}
Eq.(\ref{implement}) is satisfied if and only if  
\begin{eqnarray}
\forall \ket{\Psi} \in \mathcal{H}_{in} \otimes \mathcal{H}_{in'},\ |(\bra{\Psi}\otimes \bra{\bar{0}}) (U^{\dagger} \otimes U \otimes I) Z_x (\ket{\Psi}\otimes \ket{\bar{0}})|^2 = 1. \label{suff}
\end{eqnarray}
\end{Lemma}

\begin{Proof}
Let us show the sufficient condition, since the necessary one is trivial.
If Eq.(\ref{suff}) is satisfied, 
\begin{eqnarray}
\forall \ket{\psi} \otimes \ket{\psi'} \in \mathcal{H}_{in} \otimes \mathcal{H}_{in'}, Z_x( \ket{\psi} \otimes \ket{\psi'} \otimes \ket{\bar{0}}) = U\ket{\psi} \otimes U^{\dagger}\ket{\psi'} \otimes \ket{\bar{0}}, \\
\to U_x( \ket{\psi}_{in} \otimes \ket{\bar{0}}_{a}) \otimes \ket{\psi'}_{in'} = U\ket{\psi}_{in} \otimes U_x ( U^{\dagger}\ket{\psi'}_{in'} \otimes \ket{\bar{0}}_a ).
\end{eqnarray}
From separability, we obtain that 
\begin{eqnarray}
U_x( \ket{\psi} \otimes \ket{\bar{0}}) = U\ket{\psi} \otimes \ket{\phi_{\psi,x}}.
\end{eqnarray} 
Thus, all we have to show is that $\ket{\phi_{\psi, x}}$ does not depend on $\psi$. 
For two different inputs $\ket{\psi}$ and $\ket{\psi'}$,
\begin{eqnarray}
U_x( \ket{\psi} \otimes \ket{\bar{0}}) = U\ket{\psi} \otimes \ket{\phi_{\psi,x}}, \\
U_x( \ket{\psi'} \otimes \ket{\bar{0}}) = U\ket{\psi'} \otimes \ket{\phi_{\psi',x}}.
\end{eqnarray}
Thus, we easily derive that
\begin{eqnarray}
\braket{\psi}{\psi'}(1 - \braket{\phi_{\psi,x}}{\phi_{\psi',x}}) = 0,
\end{eqnarray}
and we can conclude that $\ket{\phi_{\psi,x}}$ does not depend on the input.
\qed
\end{Proof}

From Lemma 2, let us define exact non-identity check problem.

\begin{Definition} ({\bf Exact non-identity check})\\
Let $x \in \{ 0,1 \}^{\ast}$ be a classical description of a quantum circuit $U_x$ that acts on the Hilbert space
$
\mathcal{H}_{in} \otimes \mathcal{H}_{a} := \mathcal{B}^{\otimes n_x} \otimes \mathcal{B}^{\otimes m_x},
$
where $n_x$ and $m_x$ grow at most polynomially in $|x|$.
Then, decide whether 
\begin{eqnarray}
\exists \ket{\Psi} \in \mathcal{H}_{in} \otimes \mathcal{H}_{in'},\ |\bra{\Psi}\otimes \bra{\bar{0}} Z_x (\ket{\Psi}\otimes \ket{\bar{0}})|^2 \neq 1, \\
{\rm or}\ \forall \ket{\Psi} \in \mathcal{H}_{in} \otimes \mathcal{H}_{in'},\ |\bra{\Psi}\otimes \bra{\bar{0}} Z_x (\ket{\Psi}\otimes \ket{\bar{0}})|^2 = 1.
\end{eqnarray}
\end{Definition}

From the definition of exact non-identity check, we give exact equivalence check.

\begin{figure}
\begin{center}
\includegraphics[scale=0.7]{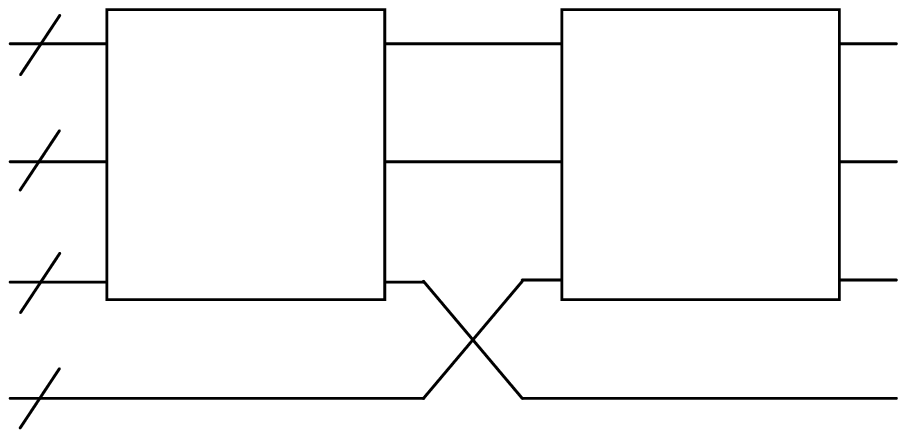}
\put(-223,78){Inputs}
\put(-223,54){Inputs$^\prime$}
\put(-228,30){Ancillas}
\put(-228,6){Ancillas$^\prime$}
\put(-184,88){$n$}
\put(-184,64){$n$}
\put(-185,40){$m_x$}
\put(-184,16){$m_y$}
\put(-147,54){\Large{$Z_x$}}
\put(-55,54){\Large{$Z_x^{\dagger}$}}
\caption{Circuit $Z_{x,y}$ consisting of $Z_x$ and $Z_y^{\dagger}$.}
\label{FigZ01}
\end{center}
\end{figure}

\begin{Definition} ({\bf Exact equivalence check})\\
Let $x$ and $y$ be classical descriptions of quantum circuits $U_x$ and $U_y$ that act on the Hilbert spaces
$
\mathcal{H}_{in} \otimes \mathcal{H}_a := \mathcal{B}^{\otimes n} \otimes \mathcal{B}^{\otimes m_x}
$ and
$
\mathcal{H}_{in'} \otimes \mathcal{H}_{a'} := \mathcal{B}^{\otimes n} \otimes \mathcal{B}^{\otimes m_y},
$
where $n$ grow at most polynomially in Max($|x|,|y|$).
Construct a quantum circuit $Z_{x,y} = [(Z_x)_{in,in',a} \otimes I_{a'}][(Z_y^{\dagger})_{in,in',a'} \otimes I_{a}]$ in Fig. \ref{FigZ01}. Then, decide whether 
\begin{eqnarray}
\exists \ket{\Psi} \in \mathcal{H}_{in} \otimes \mathcal{H}_{in'},\ |\bra{\Psi} \otimes \bra{\bar{0}} Z_{x,y} (\ket{\Psi} \otimes \ket{\bar{0}})|^2 \neq 1, \label{eq1} \\ 
{\rm or}\ \forall \ket{\Psi}  \in \mathcal{H}_{in} \otimes \mathcal{H}_{in'},\ 
|\bra{\Psi} \otimes \bra{\bar{0}} Z_{x,y} (\ket{\Psi} \otimes \ket{\bar{0}})|^2 = 1, \label{eq2}
\end{eqnarray}
where $\ket{\bar{0}} \in \mathcal{H}_{a} \otimes \mathcal{H}_{a'}$.
\end{Definition}

From the definition, exact equivalence check is clearly reduciable to exact non-identity check in polynomial time.
Let us explain the meaning of ``equivalence'' in exact equivalence check.
\begin{Lemma}
Eq.(\ref{eq2}) is satisfied if and only if $U_x$ and $U_y$ implement an identical unitary operation, {\it i.e.},
there exists a unitary operation $U$ such that 
\begin{eqnarray}
\forall \ket{\psi} \in \mathcal{H}_{in} (=\mathcal{H}_{in'}),\ U_i( \ket{\psi} \otimes \ket{\bar{0}}) = U\ket{\psi} \otimes \ket{\phi_i}, \label{equivdef}
\end{eqnarray}
for an arbitrary $i \in \{ x, y \}$.
\end{Lemma}

\begin{Proof}
We show the necessary condition, because the sufficient one is derived from direct calculation.
If Eq.(\ref{eq2}) is satisfied, for an arbitrary $\ket{\Psi} \in \mathcal{H}_{in} \otimes \mathcal{H}_{in'}$, we obtain
\begin{eqnarray}
Z_x (\ket{\Psi} \otimes \ket{\bar{0}}_a) \otimes \ket{\bar{0}}_{a'} = Z_y (\ket{\Psi} \otimes \ket{\bar{0}}_{a'}) \otimes \ket{\bar{0}}_a,
\end{eqnarray}
from which we can derive that $Z_x$ and $Z_y$ implement the identical unitary operation U, {\it i.e.},  
\begin{eqnarray}
Z_x (\ket{\Psi} \otimes \ket{\bar{0}}_a) = U\ket{\Psi} \otimes \ket{\bar{0}}_a, \label{nonsep1} \\
Z_y (\ket{\Psi} \otimes \ket{\bar{0}}_{a'}) = U\ket{\Psi} \otimes \ket{\bar{0}}_{a'}. \label{nonsep2} 
\end{eqnarray} 

Let us show that $U_x$ implements a unitary operation by contradiction.
If $U_x$ implements no unitary operation, there exists a $\ket{\psi} \in \mathcal{H}_{in}$ such that
\begin{eqnarray}
U_x (\ket{\psi} \otimes \ket{\bar{0}}_a) = \sum_{i=1}^{d>1} c_i \ket{i}_{in} \otimes \ket{i}_a,
\end{eqnarray}
where we used Schmidt decomposition and $c_i$'s are non-zero coefficients.
From Eq.(\ref{nonsep1}), for an orthonormal basis $\{ \ket{k} \}_{k=1}^{{\rm dim} \mathcal{H}_{in'}}$ of $\mathcal{H}_{in'}$,
\begin{eqnarray}
\sum_{i=1}^{d>1} c_i \ket{i}_{in} \otimes U_x^{\dagger}( \ket{k}_{in'} \otimes \ket{i}_a) = 
\sum_{i=1}^{d>1} c_i \ket{i}_{in} \otimes \ket{\phi_{ki}}_{in'} \otimes \ket{\bar{0}}_a,
\end{eqnarray}
where $\braket{\phi_{ki}}{\phi_{kj}} = \delta_{ij}$, since local unitary operation does not change entanglement.
Further, from linearity, for an arbitrary linear combination $\alpha \ket{k}_{in'} + \beta \ket{l}_{in'}, (k\neq l)$,
\begin{eqnarray}
&& U_x^{\dagger}( (\alpha \ket{k}_{in'} + \beta \ket{l}_{in'}) \otimes \ket{i}_a) = 
(\alpha \ket{\phi_{ki}}_{in'} + \beta \ket{\phi_{li}}_{in'}) \otimes \ket{\bar{0}}_a, \nonumber \\
\to && \delta_{ij} = |\alpha|^2 \braket{\phi_{ki}}{\phi_{kj}} + |\beta|^2 \braket{\phi_{li}}{\phi_{lj}} 
                    +\alpha^{\ast}\beta \braket{\phi_{ki}}{\phi_{lj}} + \alpha \beta^{\ast} \braket{\phi_{li}}{\phi_{kj}}, \nonumber \\
\to && \braket{\phi_{ki}}{\phi_{lj}} = \delta_{kl} \delta_{ij}. 
\end{eqnarray}
Since $1 \le k,l \le {\ dim} \mathcal{H}_{in'}$, $d$ is required to be one, which is the contradiction.
\qed
\end{Proof}

\section{Main Result}
Our goal in this paper is to prove the following theorem.

\begin{Theorem}
Exact non-identity check is NQP-complete.
\end{Theorem}
\begin{Proof}
First, let us show that Exact non-identity check is in NQP.
For a quantum circuit $U$ that acts on $\mathcal{H}_{in} \otimes \mathcal{H}_a = \mathcal{B}^{\otimes n} \otimes \mathcal{B}^{\otimes m}$, consider the following NQP simulation of exact non-identity check (``verifier'').

\begin{itemize}
\item For inputs $\ket{\bar{0}} \in \mathcal{H}_{in}^{\otimes 3}$, apply $H^{\otimes n}$ on the first $n$ qubits.
\item For every $i \in \{ 1,\cdots , n \}$, apply controlled-not gates on $i$th and $(n+i)$th qubit and on $i$th and $(2n+i)$th qubit Then, applying $H^{\otimes n}$ on the last $n$ qubits, we have input states
\begin{eqnarray}
\frac{1}{\sqrt{2^n}}\sum_{x=0}^{2^n-1} \ket{x}\otimes \ket{x} \otimes \ket{x_+}. \label{input}
\end{eqnarray}
\item Adding ancilla states $\ket{\bar{0}} \otimes \ket{\bar{0}} \in \mathcal{H}_a \otimes \mathcal{H}_a$ into the state in Eq.(\ref{input}), apply $I \otimes U \otimes U$ on the state. As a result, we have the state
\begin{eqnarray}
\frac{1}{\sqrt{2^n}}\sum_{x=0}^{2^n-1} \ket{x}\otimes U(\ket{x} \otimes \ket{\bar{0}}) \otimes U(\ket{x_+} \otimes \ket{\bar{0}}).
\end{eqnarray}
\item Make the measurement on the first $2n$ inputs in computational basis and on the last $n$ inputs in $\ket{x_+}$ basis. Accept if an outcome is not $(x, x, x)$.
\end{itemize}

For the completeness, we use the contradiction. 
Assume that the verifier never accepts $U$ though $U$ implements no identity.  
Remembering that a probability of an outcome $(x, y, z)$ is given by 
\begin{eqnarray}
{\rm Pr}(x,y,z) &=& \frac{1}{2^n} \bra{y} {\rm tr}_a [U(\ket{x}\bra{x} \otimes \ket{\bar{0}}\bra{\bar{0}})U^{\dagger}] \ket{y} \\ \nonumber 
&& \bra{z_+} {\rm tr}_a [U(\ket{x_+}\bra{x_+} \otimes \ket{\bar{0}}\bra{\bar{0}})U^{\dagger}] \ket{z_+}, \label{contra}
\end{eqnarray}
we can derive that ${\rm Pr}(x,x,x) = 1/2^n$ from $\sum_x {\rm Pr}(x,x,x) = 1$.
Thus, $U (\ket{x} \otimes \ket{\bar{0}})$ and $U(\ket{x_+} \otimes \ket{\bar{0}})$ are required to be 
$\ket{x} \otimes \ket{\phi_{x}}$ and $\ket{x_+} \otimes \ket{\phi'_{x}}$.
On the other hand, we can calculate directly 
\begin{eqnarray}
U(\ket{x_+} \otimes \ket{\bar{0}}) = \frac{1}{\sqrt{2^N}}\sum_y (-1)^{x\cdot y} \ket{y} \otimes \ket{\phi_y}.
\end{eqnarray}
From the separability, all the $\ket{\phi_y}$s must be equal. 
Thus, $U$ implements the identity, which contradicts the assumption. 
 
For the soundness, suppose that $U$ implements the identity. It is trivial that the verifier never accepts $U$ from the definition of the verifier.

In order to show the NQP-hardness, it is sufficient to show that QMA($\delta$, 0) is reducible to exact non-identity check, because of Lemma \ref{NQPQMA}.
Note that $\delta : \mathbb{Z}^+ \to (0,1]$.
Let $U$ be a quantum circuit of QMA($\delta$, 0) generated from $x \in \{ 0,1 \}^{\ast}$. From the definition of QMA($\delta$, 0),
at least $\delta$ completeness and perfect soundness are satisfied:
\begin{eqnarray}
(\textrm{Completeness})&& \forall x \in L \ \exists \rho \in \mathcal{S}(\mathcal{B}^{\otimes n_x}),\ {\rm Tr}[U (\rho \otimes \ket{\bar{0}}\bra{\bar{0}}) U^{\dagger}P_1] \ge \delta(|x|), \label{complete} \\
(\textrm{Soundness})&& \forall x \not\in L \ \forall \rho \in \mathcal{S}(\mathcal{B}^{\otimes n_x}),\ {\rm Tr}[U (\rho \otimes \ket{\bar{0}}\bra{\bar{0}}) U^{\dagger}P_1] = 0. \label{sound}
\end{eqnarray}

In order to apply $U$ to exact non-identity check, we construct a circuit $Z$ that implements the identity whenever there exists no state accepted by $U$ or implements no identity if there is a witness.
One qubit register is added to extend the inputs and the whole transformation is given by $Z := (X \otimes U^{\dagger})V(I\otimes U)$,
where $V$ is a controlled-$X$ which acts on the extended register and is controlled by the first qubit in the original input registers.
Note that $V$ operates when the controlled state is $\ket{0}$ (See Fig.\ref{FigZ}).
We always set the original ancillas in state $\ket{\bar{0}}$.

\begin{figure}
\begin{center}
\includegraphics[scale=0.7]{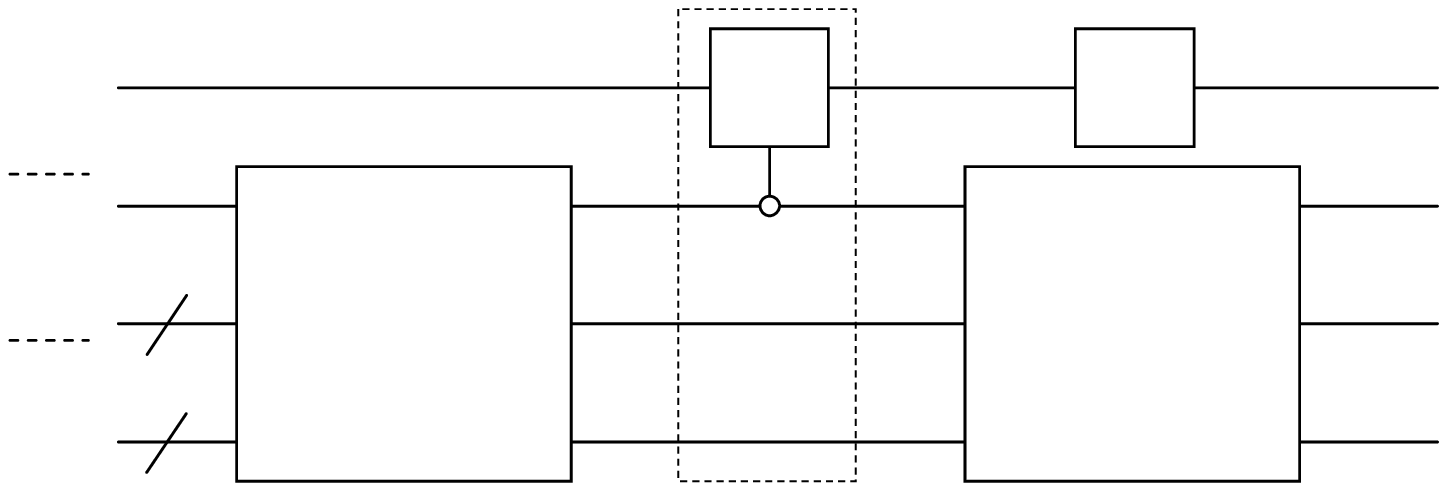}
\put(-318,96){Extended}
\put(-313,84){register}
\put(-310,55){Inputs}
\put(-315,23){Ancillas}
\put(-220,39){\Large{$U$}}
\put(-75,39){\Large{$U^{\dagger}$}}
\put(-72,90){$X$}
\put(-146,90){$X$}
\put(-146,112){\large{$V$}}
\caption{Circuit Z consisting of $U$, $U^{\dagger}$, a controlled-$X$, and $X$. Note that the controlled-$X$ operates when the controlled state is $\ket{0}$.}
\label{FigZ}
\end{center}
\end{figure}

First, let us show that when Eq.(\ref{complete}) is satisfied, $Z$ implements no identity.
Consider a proof $\ket{\phi}$ \footnote{ Note that there always exists a pure state proof accepted with $\epsilon > 0$ probability for any mixed state proof accepted with $\delta$ probability because the mixed state can be written as a probability distribution of orthogonal pure states.} that is accepted by QMA($\delta$,0)-verifier $U$ with $\epsilon >0$ probability, and apply $Z$ on $\ket{0} \otimes \ket{\phi} \otimes \ket{\bar{0}}$.
Defining that $\ket{\Psi} := \ket{0} \otimes U(\ket{\phi} \otimes \ket{\bar{0}}) = \sqrt{1-\epsilon} \ket{0}\otimes \ket{0} \otimes \ket{\Phi} + \sqrt{\epsilon} e^{i \gamma} \ket{0} \otimes \ket{1} \otimes \ket{\Phi'} $, we calculate that 

\begin{eqnarray}
Z (\ket{0} \otimes \ket{\phi} \otimes \ket{\bar{0}}) &=& (X\otimes U^{\dagger})V \ket{\Psi} \nonumber \\
&=& \sqrt{1-\epsilon} \ket{0} \otimes U^{\dagger}( \ket{0} \otimes \ket{\Phi}) \nonumber \\
&& + \sqrt{\epsilon} e^{i \gamma} \ket{1} \otimes U^{\dagger} (\ket{1} \otimes \ket{\Phi'}). \label{Eq}
\end{eqnarray}
Assuming that $Z$ implements the identity with ancilla, $Z (\ket{0} \otimes \ket{\phi} \otimes \ket{\bar{0}})$ must be a separable state in the extended register and the other systems. However, Eq.(\ref{Eq}) shows that the state is a bipartite entangled state, which contradicts the assumption.
Thus, $Z$ implements no identity for the input and extended space, {\it i.e.},
\begin{eqnarray}
\bra{0} \otimes \bra{\phi} {\rm tr}_{a} [Z( \ket{0}\bra{0} \otimes  \ket{\phi}\bra{\phi} \otimes \ket{\bar{0}}\bra{\bar{0}})Z^{\dagger}] \ket{0} \otimes \ket{\phi} < 1. 
\end{eqnarray}

Now, we show that $Z$ implements the identity if there is no witness.
It is sufficient to consider arbitrary pure states, since an arbitrary mixed state is rewritten into a probability distribution of orthogonal pure states in the extended and input Hilbert space. 
Consider a pure state $\ket{\psi} = \sum_i (c_i \ket{0} \otimes \ket{i} + d_i \ket{1} \otimes \ket{i})$ in the extended and input Hilbert space, 
where $\sum_i (|c_i|^2 + |d_i|^2) = 1, c_i, d_i \in \mathbb{C}$. 
Applying $Z$ on $\ket{\psi} \otimes \ket{\bar{0}}$, we derive
\begin{eqnarray}
Z(\ket{\psi} \otimes \ket{\bar{0}}) &=& (X \otimes U^{\dagger})V\sum_i \big( c_i \ket{0} \otimes U(\ket{i} \otimes \ket{\bar{0}}) + d_i \ket{1} \otimes U(\ket{i} \otimes \ket{\bar{0}}) \big) \nonumber \\
&=& (X \otimes U^{\dagger})\sum_i \big( c_i \ket{1} \otimes U(\ket{i} \otimes \ket{\bar{0}}) + d_i \ket{0} \otimes U(\ket{i} \otimes \ket{\bar{0}}) \big) \nonumber \\
&=& \ket{\psi} \otimes \ket{\bar{0}},
\end{eqnarray}
where we used the assumption that $U$ accepts no input state, \textit{i.e.}, Eq.(\ref{sound}). 
Therefore, we conclude that for an arbitrary quantum state $\ket{\psi}$,
\begin{eqnarray}
\bra{\psi} {\rm tr}_{a} [Z(\ket{\psi}\bra{\psi} \otimes \ket{\bar{0}}\bra{\bar{0}})Z^{\dagger}] \ket{\psi} = 1. 
\end{eqnarray}
\qed
\end{Proof}

From the NQP-completeness of exact non-identity check, we derive that exact equivalence check is NQP-complete.
\begin{Corollary}
Exact equivalence check is NQP-complete.
\end{Corollary}

\begin{Proof}
Reduction from exact equivalence check to exact non-identity check is trivial.
To reduce exact non-identity check to exact equivalence check, for a given $U_x$, take $U_y = I$ in Fig. 2. \qed
\end{Proof}

\section{Quantum gates minimization problem}

As an application of exact non-identity check, we introduce a quantum gates minimization problem of minimizing quantum resources of a quantum gate array without changing the implemented unitary operation, and show that this problem is NQP-hard.

We prepare definitions for quantum gates minimization problem.
$|U_x|$ denotes the number of quantum gates constructing a quantum circuit $U_x$ generated by a classical inputs $x \in \{ 0,1 \}^{\ast}$ with respect to fixed universal quantum gates. 
Define 
\begin{eqnarray}
S_U = \{ x\in \{ 0,1 \}^{\ast}\ |\ \exists \ket{\phi_i} \in \mathcal{H}_{a}\ \forall \ket{\psi} \in \mathcal{H}_{in},\ U_x( \ket{\psi} \otimes \ket{\bar{0}}) = U\ket{\psi} \otimes \ket{\phi_i} \}
\end{eqnarray} 
as a set of equivalent classical descriptions of an implemented unitary operation $U$. 
\begin{Definition} ({\bf Quantum gates minimization problem})\\
A classical description $x$ for a quantum circuit $U$ is said to be minimized if $|U_x| = \min_{y \in S_{U}} |U_y|$.
For a given classical description $x$, quantum gates minimization is said to be feasible if a minimized classical description of $x$ is computed in polynomial-time of $|x|$.  
\end{Definition}

So far, we have a corollary from the NQP-completeness of exact non-identity check.

\begin{Corollary}
Quantum gates minimization problem is NQP-hard.
\end{Corollary}

\begin{Proof}
For a given classical description $x$, when $U_x$ implements the identity $U=I$, 
\begin{eqnarray}
\min_{y \in S_{U}} |U_y| = 0.
\end{eqnarray}
When $U_x$ implements no identity, $\min_{y \in S_{U_x}} |U_y|  > 0$. \qed
\end{Proof}

\section{Summary}

We defined exact non-identity check problem of deciding whether a given classical description of a quantum circuit is strictly equivalent to the identity or not, and showed that this problem is NQP-complete.
Exact non-identity check is a decision problem of quantum circuits and is useful to analyze quantum gate complexity. 
For example, as corollaries of our result, we proposed exact equivalence check and quantum gate minimization problem and showed the NQP-completeness and the NQP-hardness respectively.
 
\section{Acknowledgements}
The author thanks M. Murao, M. Ukita and Y. Kawamoto for useful discussions.


\begin{thebibliography}{0}

\bibitem{identity} D. Janzing, P. Wocjan, and T. Beth, {\it Int. J. Quantum Inf.} {\bf 3} (2005) 463.

\bibitem{NQPDef} L. Adleman, J. DeMarrais, and M. Huang, Quantum computability, {\it SIAM Journal on Computing} {\bf 26} (1997) 1524-1540. 

\bibitem{NQMA} H. Kobayashi, K. Matsumoto, and T. Yamakami,
Quantum Merlin-Arthur Proof Systems: Are Multiple Merlins More Helpful to Arthur?, quant-ph/0306051v2

\bibitem{Kitaev} A. Y. Kitaev, A. H. Shen, and M. N. Vyalyi,
{\it Classical and Quantum Computation}, Graduate Studies in Mathematics, Vol 47 (American Mathematical Society, 2002).

\bibitem{NQP} S. Fenner, F. Green, S. Homer, and R. Pruim,
Determining Acceptance Possibility for a Quantum Computation is Hard for the Polynomial Hierarchy, quant-ph/9812056v1.

\end{thebibliography}
\end{document}